\documentclass[11pt]{iopart} %
\usepackage{iopams}%

\usepackage[utf8]{inputenc}%
\usepackage{color}%
\usepackage{graphicx}%

\newcommand{\Drs}{\delta r^2}%
\newcommand{\Drsf}{\delta r^2_\mathrm{free}}%
\newcommand{\Drst}{\delta r^2_\mathrm{trap}}%
\newcommand{\Fs}{F_\mathrm{s}}%
\newcommand{\Fsf}{F_\mathrm{s}^\mathrm{free}}%
\newcommand{\Fst}{F_\mathrm{s}^\mathrm{trap}}%
\newcommand{\Gs}{G_\mathrm{s}}%
\newcommand{\Gsf}{G_\mathrm{s}^\mathrm{free}}%
\newcommand{\Gst}{G_\mathrm{s}^\mathrm{trap}}%
\newcommand{\Nf}{N_\mathrm{f}}%
\newcommand{\Nff}{N_\mathrm{f}^\mathrm{free}}%
\newcommand{\Nft}{N_\mathrm{f}^\mathrm{trap}}%
\newcommand{\Nm}{N_\mathrm{m}}%
\newcommand{\Phif}{\phi_\mathrm{f}}%
\newcommand{\Phim}{\phi_\mathrm{m}}%
\newcommand{\Zf}{z_\mathrm{free}}%
\newcommand{\Zt}{z_\mathrm{trap}}%

\newcommand{\Cf}{cf.}%
\newcommand{\Ie}{i.e.}%

\begin{document}%

\title[Disentangling crowding and confinement]{Dynamic arrest of colloids
in porous environments: disentangling crowding and confinement}%

\author{Jan Kurzidim$^1$, Daniele Coslovich$^2$, and Gerhard Kahl$^1$}%

\address{$^1$Institut für Theoretische Physik and Center for Computational
Materials Science (CMS), Technische Universität Wien, Wien, Austria \\
$^2$Laboratoire Charles Coulomb UMR 5221, Université Montpellier 2 and CNRS,
Montpellier, France}%

\ead{kurzidim@cmt.tuwien.ac.at}%

Date: \today

\begin{abstract}
Using numerical simulations we study the slow dynamics of a
colloidal hard-sphere fluid adsorbed in a matrix of disordered
hard-sphere obstacles. We calculate separately the contributions to
the single-particle dynamic correlation functions due to free and
trapped particles. The separation is based on a Delaunay
tessellation to partition the space accessible to the centres of
fluid particles into percolating and disconnected voids. We find
that the trapping of particles into disconnected voids of the matrix
is responsible for the appearance of a nonzero long-time plateau in
the single-particle intermediate scattering functions of the full
fluid. The subdiffusive exponent $z$, obtained from the logarithmic
derivative of the mean-squared displacement, is observed to be
essentially unaffected by the motion of trapped particles: close to
the percolation transition, we determined $z \simeq 0.5$ for both
the full fluid and the particles moving in the percolating void.
Notably, the same value of $z$ is found in single-file diffusion and
is also predicted by mode-coupling theory along the
diffusion-localisation line. We also reveal subtle effects of
dynamic heterogeneity in both the free and the trapped component of
the fluid particles, and discuss microscopic mechanisms that
contribute to this phenomenon.
\end{abstract}%

\submitto{\JPCM}%

\maketitle

\section{Introduction}%
\label{sec:Introduction}%

As colloidal fluid particles move within the external, random field
induced by a quenched, disordered arrangement of particles
(``matrix''), they can be either hindered in their motion by so-called
cages formed by other fluid (mobile) particles, or trapped by the
immobile particles of the matrix. In the former case, these cages
eventually open due to the steady, collective motion of the fluid,
while in the latter case whether or not a fluid particle can escape a
certain region of space depends only on pre-set geometric and/or
energetic restrictions. Each of these processes alone as well as their
interplay are responsible for slowing down the dynamics of the fluid,
leading thereby to a complex dynamic behaviour of the system.

These observations represent yet another confirmation that the
properties of a fluid that is exposed to a random porous matrix differ
substantially from those of a bulk fluid. While these phenomena have
been extensively studied in experiment for a wide variety of systems
(e.g., \cite{gelb1999, rosinberg1999, mckenna2000, mckenna2003,
alcoutlabi2005, mckenna2007}), related activities in theory and
computer simulations considerably lag behind. This discrepancy is
mainly due to two problems: (i) a realistic parametrisation of typical
matrices studied in experiment is very difficult, and (ii) the
evaluation of the physical properties of the fluid particles is
conceptually difficult and numerically expensive. As a consequence, so
far most, of the theoretical and simulation-based studies have focused
on \textit{static} structural \cite{lomba1993, vega1993,
meroni1996, paschinger2000}, and thermodynamic properties \cite{rosinberg1994,
kierlik1995, kierlik1997, alvarez1999, paschinger2000,
paschinger2001}. Only during the past few years, investigations on
the \textit{dynamic} properties of such systems have come within
reach \cite{gallo2003, kim2003, chang2004, mittal2006}.

Recently, a major breakthrough has been achieved in the realm of
theory: using concepts from the statistical description of so-called
``quenched-annealed'' (QA) mixtures (see below), a framework has been
put forward that allows to determine dynamic correlation functions of
fluids confined in a porous matrix \cite{krakoviack2005,
krakoviack2007, krakoviack2009}. The theory represents an extension
to mode-coupling theory (MCT) \cite{gotze1992}, a successful
framework that can be used to obtain the \textit{dynamic}
correlation functions of a system solely from knowledge about its
\textit{static} structure functions. We shall henceforth refer to
the theory as ``ROZ-MCT'', owing to the fact that it is based on
mathematical methods used in deriving the replica Ornstein-Zernike
(ROZ) equations \cite{given1994}, and that it can use structural
information provided by the ROZ equations as a source of input.

Meanwhile, thanks to the considerably increased computational power of
present-day computers, it is possible to systematically investigate
the slow dynamic properties of fluids confined in a disordered
environment in computer models. Simulating such systems is costly for
multiple reasons: (i) long simulations are necessary to reveal the
dynamic slowing-down of dense systems (ii) a large body of system
parameter combinations needs to be considered (iii) many independent
realisations of the same system have to be generated. The latter
requirement arises from the prescription to evaluate the physical
properties of QA systems, which involves an explicit double average
procedure: the first average is the usual thermodynamic one to be
taken over all possible configurations of fluid particles given a
fixed matrix configuration, the second average is a disorder average
to be performed over all possible matrix representations fulfilling
pre-imposed criteria.

To date, the ROZ-MCT framework has been applied only to QA models in
which hard-sphere particles are confined in a hard-sphere
matrix---either with fluid and matrix being of equal size \cite{krakoviack2005,
krakoviack2007, krakoviack2009} or different sizes
\cite{kurzidim2009}. Within the QA model, a matrix configuration is
obtained by taking a snapshot of an equilibrated fluid (quenched
component); subsequently the particles of the fluid (annealed
component) are immersed in this configuration. Shortly after
publication of the theoretical predictions, extensive and systematic
molecular dynamics (MD) simulations have been performed for precisely
this system \cite{kurzidim2009, kurzidim2010, kim2009}. While
simulations confirmed most of the intriguingly complex features that
the theory predicted for the kinetic diagram (such as a discontinuous
and a continuous glass transition as well as a continuous
diffusion-localisation transition), others are yet to be found (in
particular the re-entrant glass transition scenario). In this context,
it is worth mentioning other recent works which investigate the
dynamics of a fluid \cite{chang2004, mittal2006, gallo2009} or of
tracer particles \cite{hoefling2006, hoefling2007, babu2008,
sung2008} moving in a matrix of quenched particles.

Our previous work \cite{kurzidim2009, kurzidim2010} was dedicated
to an ``overall'' investigation of the dynamic properties (in terms of
both single-particle and collective dynamic properties) of hard
spheres confined in a disordered matrix of hard-sphere particles, and
to a comparison of these results to the theoretical predictions \cite
{krakoviack2005, krakoviack2007, krakoviack2009}. Computer
simulations offer the distinct possibility to proceed one step
further: in contrast to the theoretical framework, in simulations it
is possible to distinguish between caging and trapping by
geometrically analysing the voids formed by the matrix particles.

The pertinent procedure reads as follows. Given a particular
realisation of a state point, the first step is to classify the voids
formed by the matrix particles while disregarding the fluid particles.
The algorithm used is detailed in Ref. \cite{kurzidim2011a}; in
short, we distinguish between \textit{disconnected} voids of finite
volume from which---due to geometric restrictions---particles cannot
escape, and the \textit{percolating} void that---taking into
account the periodic boundary conditions---extends over the entire
space and in which particles can propagate infinitely far. (Note that
multiple percolating voids can only exist in systems of finite size.)
In the second step, for each fluid particle in the system realisation
we use this void analysis to determine in which class of void the
particle is located. The fluid particles thus fall in two classes:
``trapped'' particles (superscript ``trap'') that populate the
disconnected voids, and ``free'' particles (superscript ``free'') that
move in the percolating void. We then compute a set of single-particle
dynamic correlation functions [namely the self-intermediate scattering
function $\Fs(k,t)$, the mean squared displacement $\Drs(t)$, and the
self-part of the direction-averaged van Hove function $\Gs(r,t)$]
separately for the trapped and the free particles [\Ie, $\Fst(k,t)$
and $\Fsf(k,t)$, etc.].

In an effort to specify and to quantify the respective contributions
of trapped and free particles to the dynamic structure, $\Fs(k,t)$ and
$\Drs(t)$ are discussed along two specific pathways in the system's
parameter space. The latter is spanned by the packing fraction of the
fluid, $\Phif$, and that of the matrix, $\Phim$. The properties of
$\Gs(r,t)$ for the trapped and the free particles are discussed for
selected state points in the $(\Phim,\Phif)$ plane.

Notwithstanding the presentation of \textit{quantitative} results
in Sec.~\ref{sec:Results}, it is worth clarifying what \textit{qualitative}
behaviour should be expected for the dynamic correlation
functions. In the long-time limit, the contributions of the trapped
and the free particles differ markedly, which bears physical
significance. While the single-particle intermediate scattering
function of the free particles will always relax completely for $t
\rightarrow \infty$, this is never the case for $\Fst(k,t)$. On the
other hand, the mean-squared displacement of the free particles will
always recover diffusive behaviour while $\Drst(t)$ will always
saturate as $t \rightarrow \infty$, with the limiting value reflecting
the average pore size. Finally, for the free particles the self-part
of the van Hove function will rapidly relax towards zero for distances
in the vicinity of zero. Conversely, $\Gst(r{\simeq}0,t)$ will \textit
{not} decay below a value significantly above zero. Instead, it will
approach a steady-state distribution that reflects the distribution in
trap sizes and shapes.

The paper is organised as follows: in the subsequent section
(Sec.~\ref{sec:Model.and.theoretical.approaches}) we briefly
summarise the model and our simulation technique (referencing previous
publications for more details). Section~\ref{sec:Results} is
dedicated to the results: after a brief summary of our void
identification algorithm as well as a short discussion of the void
analysis of the state points treated in this contribution, we present
in detail our results for the single-particle correlation functions.
In Sec.~\ref{sec:Discussion} we discuss implications of those
results, and present explanations for the observations. The paper is
closed with concluding remarks.

\section{Model and theoretical approaches}%
\label{sec:Model.and.theoretical.approaches}%

Following our previous work, we have used the QA protocol to describe
the properties of fluid particles that are exposed to an external
field of a random porous matrix. In this picture, the fluid particles
move within the space left free by the matrix: both types of particles
interact with each other, but the mobile fluid particles cannot
displace the matrix particles. Choosing all particles to interact as
hard spheres, any state point is fully characterised by a combination
of $\Phim$ and $\Phif$. Different, but equivalent matrix
configurations of a given state point are obtained from an
equilibrated hard-sphere fluid at a prescribed $\Phim$. For this, a
pertinent simulation is halted at different times and the particle
positions are subsequently used as positions of the matrix particles
in a QA system. In all MD simulations we have used an event-driven MD
algorithm adapted to keeping the matrix particles at fixed positions.
Periodic boundary conditions and the minimum image convention were
employed.

As prescribed by thermodynamics \cite{given1992, given1992a,
given1994}, observables are obtained from a double-averaging
procedure: the first trace, taken over the degrees of freedom of the
fluid particles for a particular matrix configuration, is realised
using a time average along the simulation run; the second one, taken
over different yet equivalent matrix configurations, is realised as an
ensemble average over independent simulation runs. All state points
were studied via ensembles in which the number of fluid particles
($\Nf$) and the number of matrix particles ($\Nm$) added up to at
least 1000. For systems with elevated matrix packing fractions this
number was considerably enhanced in an effort to guarantee that for
all state points considered the ensemble contain at least 50 fluid
particles. A method for finding an initial configuration of QA systems
at large values of the total packing fraction, $\Phim + \Phif$, has
been presented and discussed in the Appendix of Ref.~\cite{kurzidim2010}.

In the first phase of the simulation the system is equilibrated; for
clarifying comments on this delicate issue we refer the reader to
Section II of Ref.~\cite{kurzidim2010}. In case the system was
successfully brought to equilibrium, we subsequently recorded the
particle positions in a production run extending over the same
simulation time as the equilibration run. We typically allowed for a
maximum of $30\,000\,\tau$ for the equilibration run; for selected
systems this figure was increased to the tenfold value. Here $\tau =
\sqrt{m \sigma^2 / k_\mathrm{B} T}$ is the unit of time \cite{allen1987},
where the particle mass $m$, the temperature $T$, and the
particle diameter $\sigma$ were chosen to be unity.

The definitions of all observed quantities presented in this work
[$\Fs(k,t)$, $\Drs(t)$, $z(t)$, and $\Gs(r,t)$] have been compiled in
Ref.~\cite{kurzidim2010}, to which we refer the reader.

\section{Results}%
\label{sec:Results}%

\subsection{Void analysis}%
\label{subsec:Void.analysis}%

Due to the statistic nature of our matrix model, at any value of the
matrix packing fraction $\Phim$ there is a nonvanishing probability
for the immobile matrix particles to form traps -- void spaces from
which fluid particles enclosed therein are not able to escape due to
geometric restrictions. At low values of $\Phim$ the number of traps
is small, and most fluid particles can move infinitely far away from
their initial location since the void in which they are located
extends (taking into account the periodic boundary conditions) over
the entire space. The latter volume, if present, is termed the
``percolating'' void. At sufficiently high $\Phim$, on the other hand,
all voids will be of finite size and represent traps from which the
fluid particles cannot escape. The transition between the two
scenarios occurs at the so-called percolation transition. (For an
overview over the large field of percolation transitions see, e.g.,
\cite{stauffer1995}). For an infinitely-large system this
transition is sharp, \Ie, it occurs at some well-defined value
$\Phim^*$ of the matrix packing fraction. However, if the system is
represented by a finite number of particles then this transition is
smeared out; for this reason in computer simulations matrix
realisations featuring a percolating void are encountered also beyond
$\Phim^*$.

\begin{figure}%
\begin{tabular}{c @{\hspace{1.0cm}} c}%
\includegraphics[width = 0.45\textwidth]{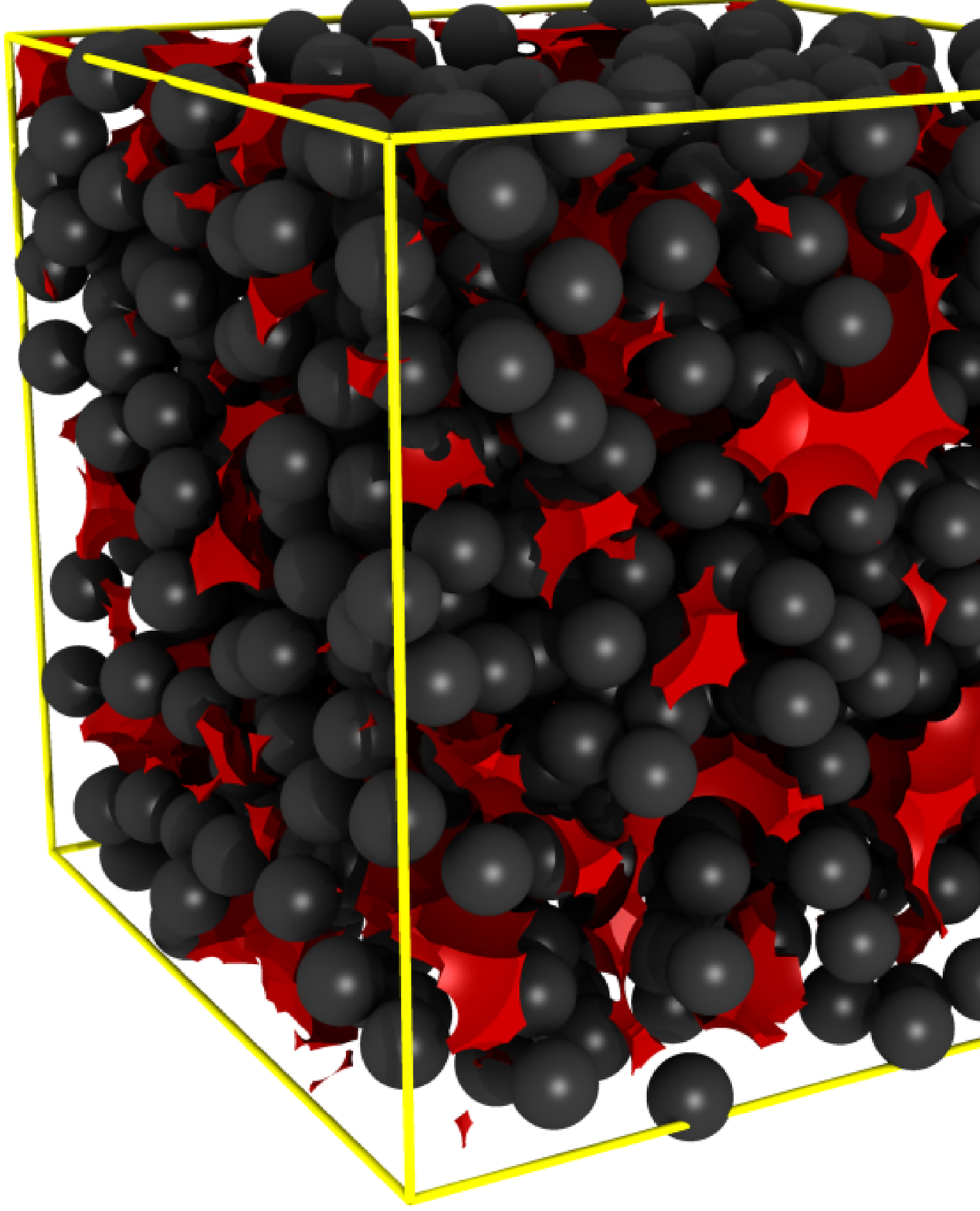}
&
\includegraphics[width = 0.45\textwidth]{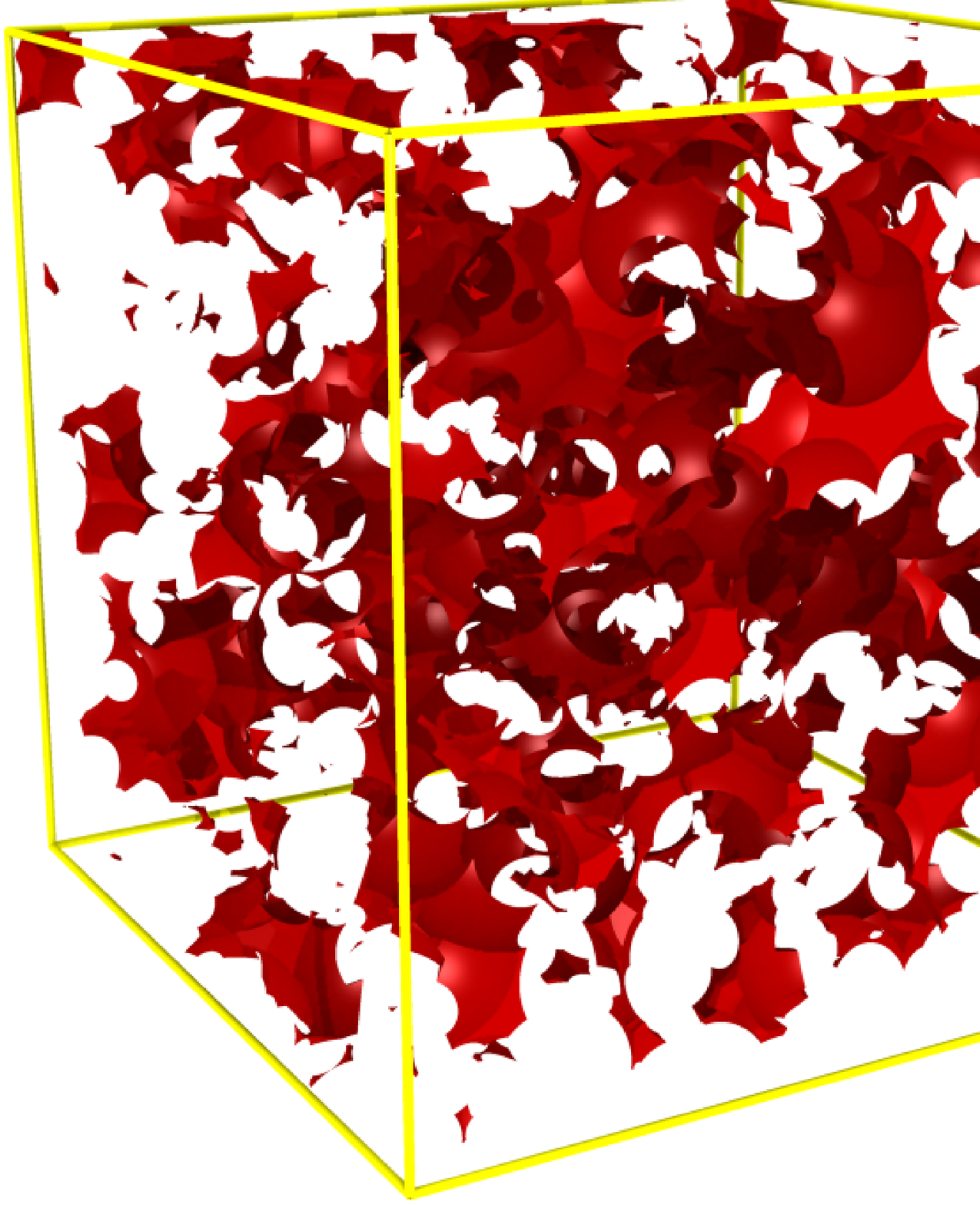}
\\ (a) & (b) \vspace{1.0cm} \\
\includegraphics[width = 0.45\textwidth]{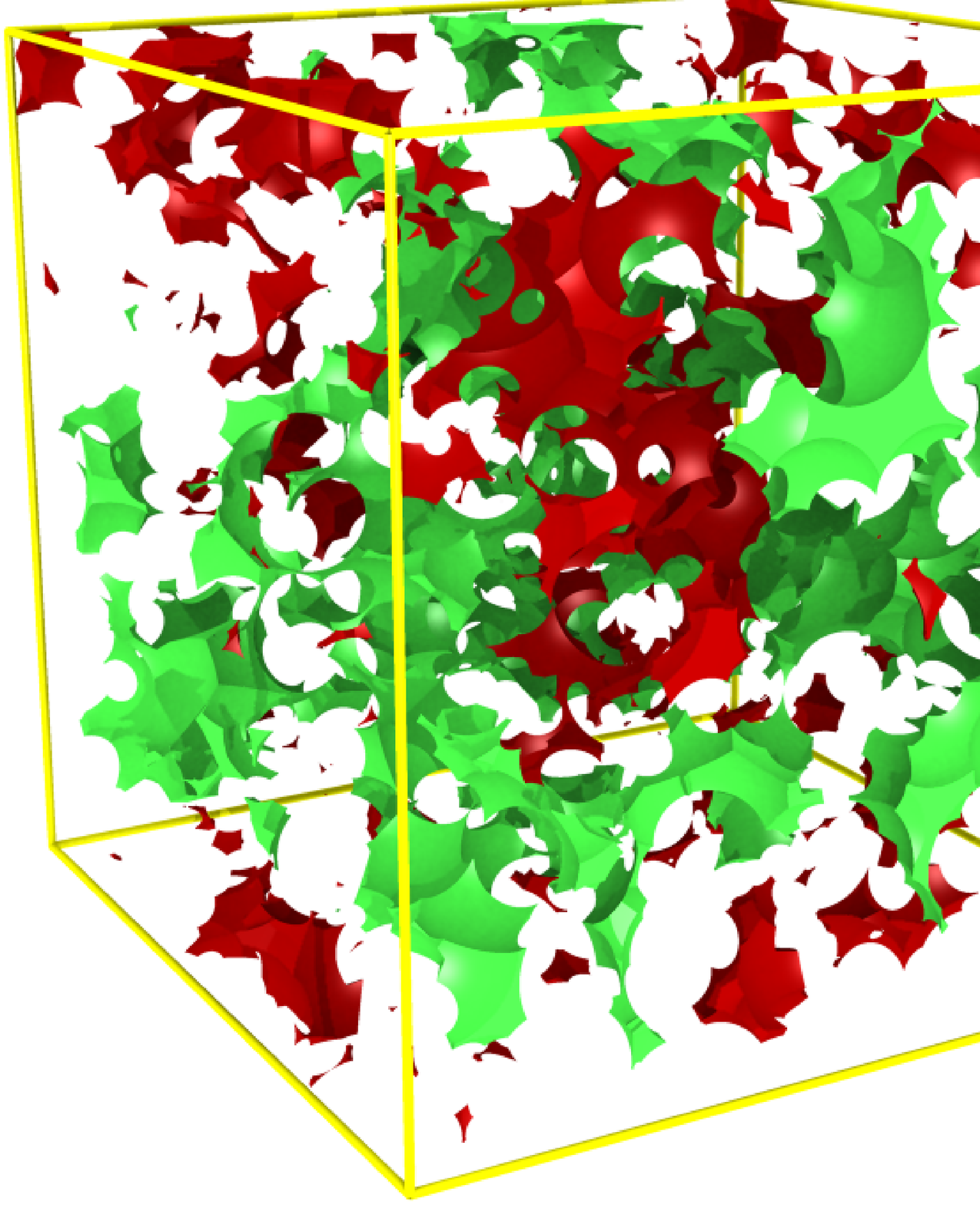}
&
\includegraphics[width = 0.45\textwidth]{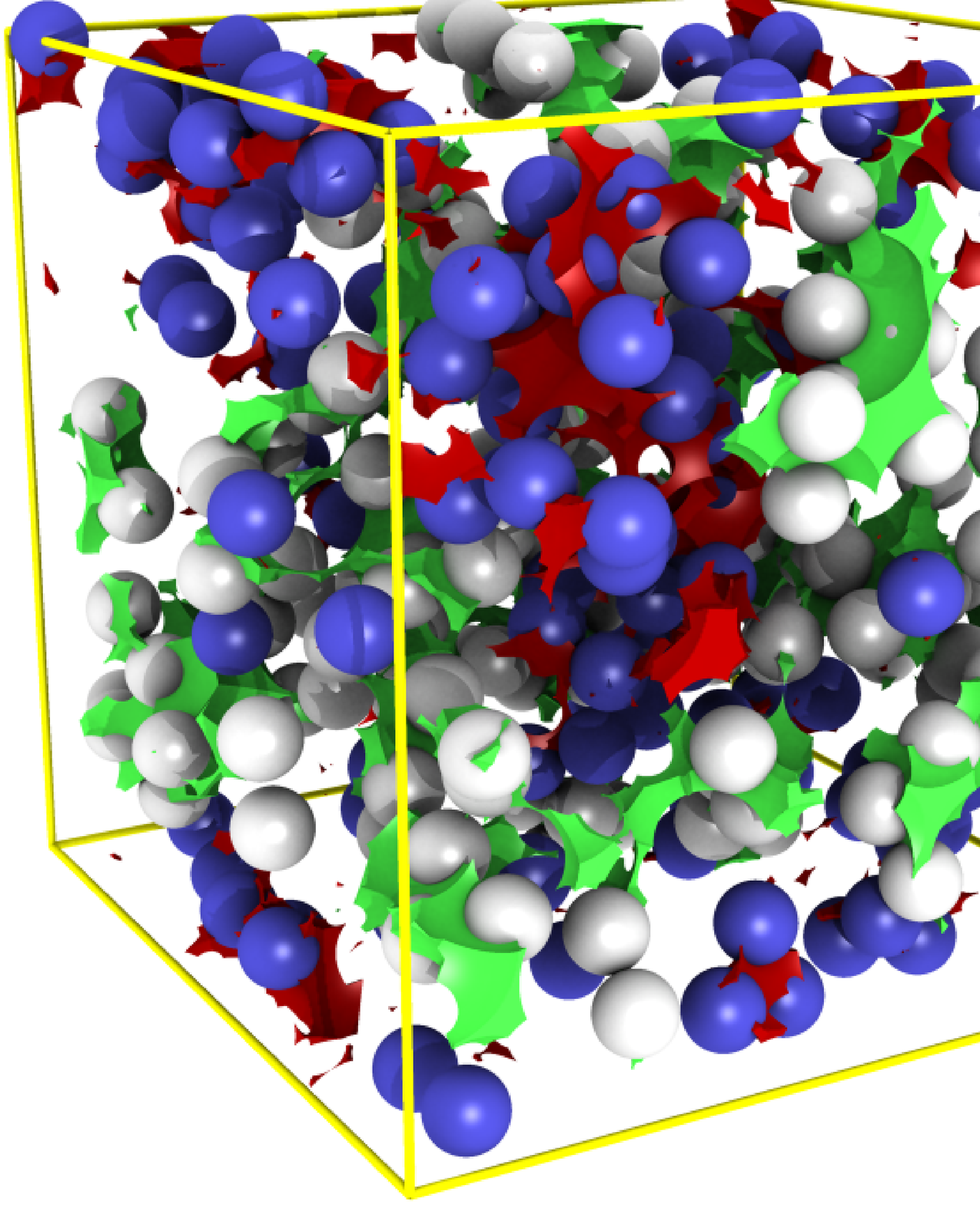}
\\ (c) & (d)%
\end{tabular}%
\caption{\label{fig:separation}%
Visualisation of the procedure to identify trapped and free fluid
particles. (a) Dark grey spheres: matrix particles. Volumes in
lighter (red) shade: voids (volumes accessible to the fluid
particles). (b) Same as previous, with the matrix particles
removed. (c) Same as previous, with the percolating void
highlighted in a light (green) shade. (d) Same as previous, with a
sample set of fluid particles inserted into the voids. Light
(grey) spheres: free particles. Dark (blue) spheres: trapped
particles.}%
\end{figure}%

We employed an algorithm based on a Delaunay tessellation---details
are presented in Ref.~\cite{kurzidim2011a}---in order to identify
for a given matrix configuration whether a void constitutes a trap or
the percolating void. Using the position of a fluid particle at an
arbitrary instance of time, it is thus possible to determine whether
the particle in question is ``free'' or ``trapped'' (according to the
previously-introduced notion). The method, which has recently received
considerable attention in contexts similar to the one of this work
\cite{sung2008, spanner2011, kim2011}, is sketched in Fig. \ref{fig:separation}.

\subsection{Dynamic correlation functions}%
\label{subsec:Dynamic.correlation.functions}%

Making use of the distinction in trapped and free particles, the
single-particle correlation functions $\Fs(k,t)$, $\Drs(t)$, and
$\Gs(r,t)$ can be split up in an additive way into two
component-weighted contributions: one stemming from the $\Nft$ trapped
particles (index ``trap''), the other one from the $\Nff$ free
particles (index ``free''). Since the fraction $\Nft/(\Nff+\Nft)$
increases with $\Phim$, the correlation functions of the trapped
particles will suffer from poor statistics low $\Phim$, whereas for
$\Phim > \Phim^*$ the results for the correlation functions of the
free particles will be affected by a relatively large numerical error.

\begin{figure} \centering
\includegraphics[width = 0.7\textwidth]{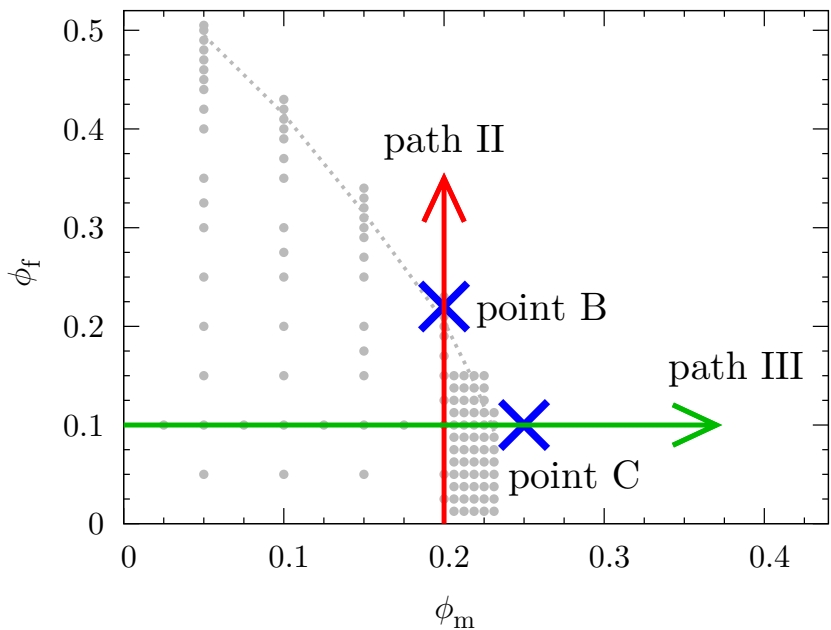}%
\caption{\label{fig:kin_diagr}%
Parameter space for QA systems of hard spheres, spanned by the
packing fraction of the matrix particles, $\Phim$, and that of the
fluid particles, $\Phif$. The paths and points relevant to this
contribution are indicated as large arrows and crosses. For
reference, they are superimposed to data taken from Fig.~4 in
Ref.~\cite{kurzidim2010}, in which simulations were performed
at state points indicated by the solid circles, and systems right
of the dashed line were defined to be ``arrested'' based on
$\Fs(k{=}7,t)$.}%
\end{figure}%

To facilitate comparison with previous data \cite{kurzidim2009,
kurzidim2010}, in this contribution we discuss the results in a
similar fashion. This includes the wave vector at which we examine
$\Fs(k,t)$, namely $k = 7$ [this value is close to the first peak in
the structure factor $S(k)$], as well as the choice of specific paths
in the parameter space---depicted in Fig.~\ref{fig:kin_diagr}---at
which we evaluated $\Fs(k,t)$ and $\Drs(t)$. Path~I in these works
(extending at fixed, low $\Phim$ while increasing $\Phif$) will not be
considered here since trapping---the key property studied in this
contribution---plays only a minor role in this region of the kinetic
diagram. Path~II denotes systems at the intermediate fixed value
$\Phim = 0.2$ and increasing fluid packing fractions; note that the
matrix packing fraction along path II is lower than $\Phim^*$ \cite{kurzidim2009,
kurzidim2010, kurzidim2011a}. Finally, along path~III
$\Phif$ is kept fixed at $0.1$ while increasing $\Phif$ over an
interval that includes the percolation transition of the voids at
$\Phim^*$.

\begin{figure} \centering
\includegraphics[width = 0.7\textwidth]{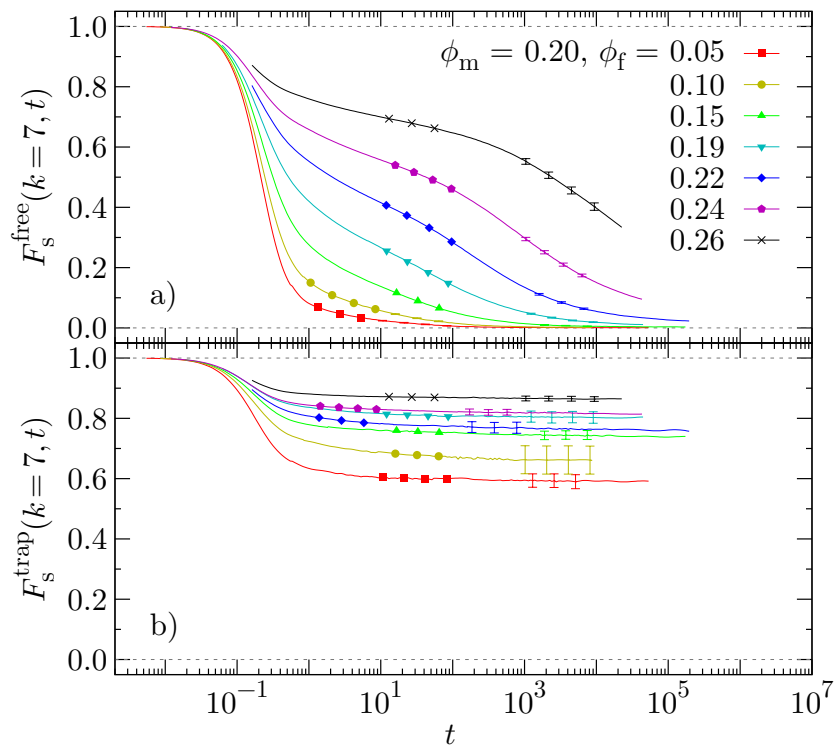}%
\caption{\label{fig:path2_fs}%
Single-particle intermediate scattering function as a function of
time $t$ at fixed $k=7$ and $\Phim = 0.20$ for a series of
$\Phif$. (a) for the free particles, $\Fsf(k,t)$, (b) for the
trapped particles, $\Fst(k,t)$. Error bars represent one standard
deviation of the mean for different system realisations.}%
\end{figure}%
\begin{figure} \centering
\includegraphics[width = 0.7\textwidth]{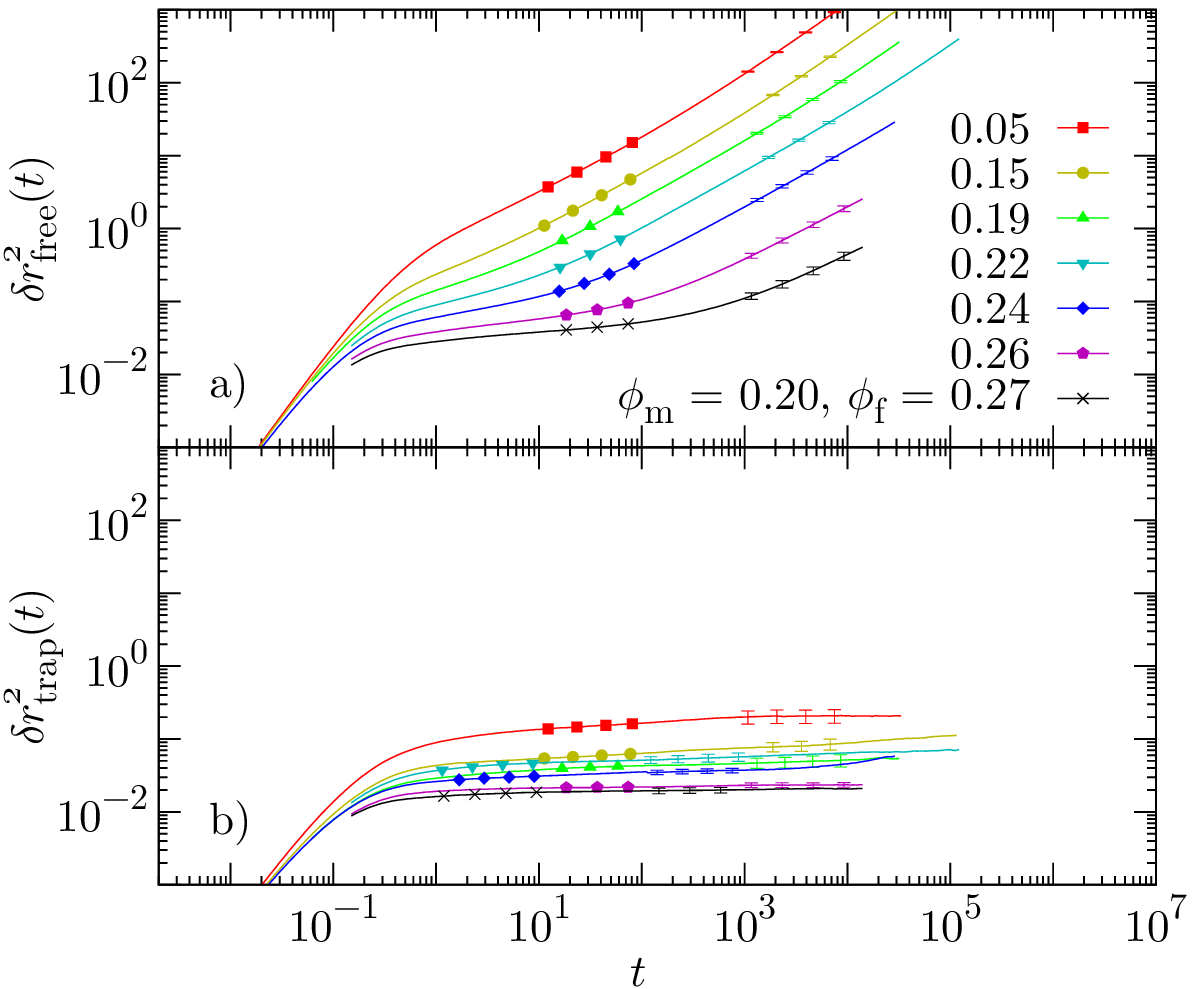}%
\caption{\label{fig:path2_msd}%
Mean-squared displacement as a function of time $t$ at fixed
$\Phim = 0.20$ for a series of $\Phif$. (a) for the free
particles, $\Drsf(t)$, (b) for the trapped particles, $\Drst(t)$.
Error bars: see Fig.~\ref{fig:path2_fs}.}%
\end{figure}%
\begin{figure} \centering
\includegraphics[width = 0.7\textwidth]{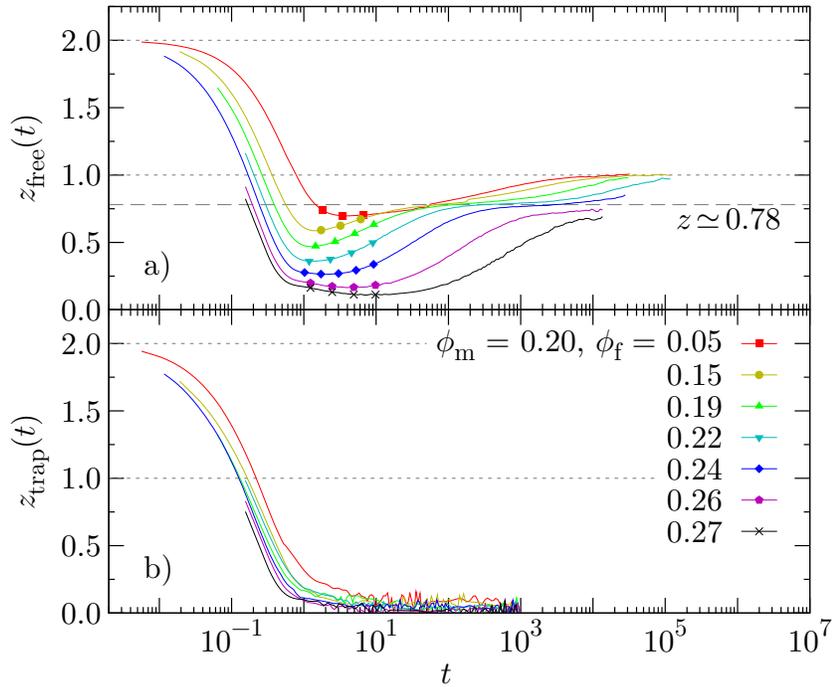}%
\caption{\label{fig:path2_z}%
Logarithmic derivative of the mean-squared displacement as a
function of time $t$ at fixed $\Phim = 0.20$ for a series of
$\Phif$. (a) for the free particles, $\Zf(t)$, (b) for the trapped
particles, $\Zt(t)$.}%
\end{figure}%

We start with the discussion of $\Fs(k,t)$ along path~II; the
respective contributions of the free and the trapped particles are
displayed in Fig.~\ref{fig:path2_fs}. Essentially, $\Fsf(k{=}7,t)$
exhibits the same behaviour as the total correlator, $\Fs(k{=}7,t)$
(shown in Fig.~8(b) of \cite{kurzidim2010}), except for the fact
that the long-time tail now tends towards \textit{zero} for all
equilibrated samples. In contrast, $\Fst(k{=}7,t)$ shows a long-time
plateau that increases with $\Phif$, which can be explained as
follows: since we keep the value of $\Phim$ fixed, the void size
distribution is the same for all systems along path~II. The increase
in the plateau is therefore due to the increasing number of trapped
fluid particles, which cause mutual jamming effects inside the
trapping voids that in turn prevent the correlator from fully relaxing
as $t$ tends to infinity.

The consequences of an increasing fluid packing fraction are even more
obvious in the respective contributions to the mean-squared
displacement, $\Drs(t)$, displayed in Fig.~\ref{fig:path2_msd}. In
the long-time limit, $\Drsf(t)$ shows normal diffusive behaviour for
all equilibrated systems. In contrast, in the long-time limit of
$\Drst(t)$ a saturation is observed; the plateau values decrease as
$\Phif$ is increased, which is due to the crowding to which the fluid
particles are subjected for larger fluid packing fractions. Assuming
an instantaneous power-law behaviour $\Drs(t) \propto t^z$, the
respective long-time behaviour of both contributions to the
mean-squared displacement is nicely reflected in the subdiffusion
exponent $z(t)$, as displayed in Fig.~\ref{fig:path2_z}: it quickly
drops to zero for the trapped particles, whereas for the free
particles $\Zf(t)$ tends to unity as $t \rightarrow \infty$, which
corresponds to diffusive behaviour (for the three lowermost curves in
Fig.~\ref{fig:path2_z}(a) the limit is not approached yet).
Notably, for several decades in time an intermediate-time plateau can
be observed in $\Zf(t)$; the value of this plateau is virtually the
same as that found for the exponent pertaining to $\Drs(t)$ for the
full fluid (see Fig.~9(b) in \cite{kurzidim2010}).

\begin{figure} \centering
\includegraphics[width = 0.7\textwidth]{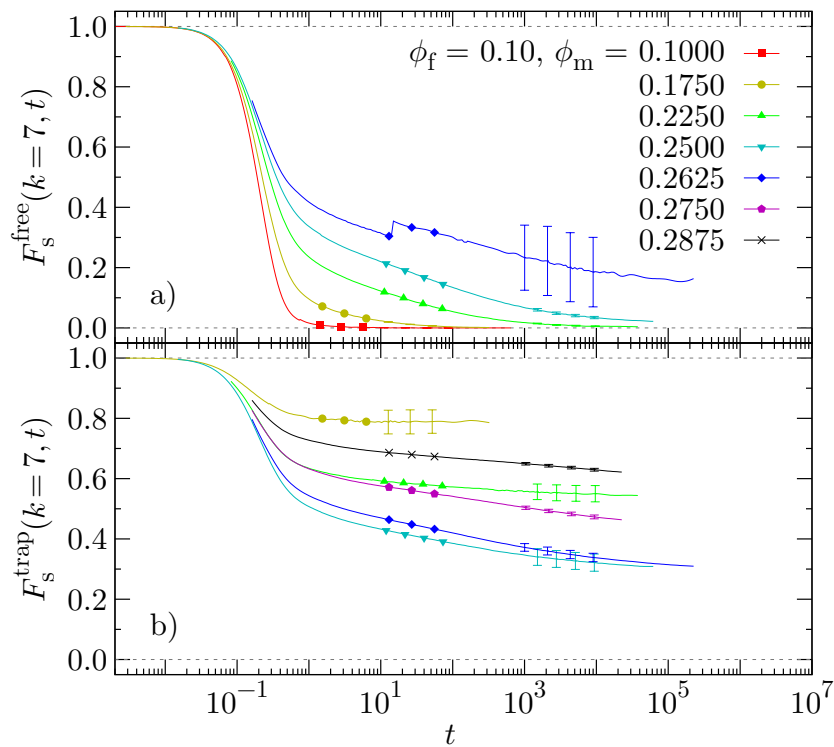}%
\caption{\label{fig:path3_fs}%
Single-particle intermediate scattering function as a function of
time $t$ at fixed $k=7$ and $\Phif = 0.10$ for a series of
$\Phim$. (a) for the free particles, $\Fsf(k,t)$, (b) for the
trapped particles, $\Fst(k,t)$. Error bars: see Fig.~\ref{fig:path2_fs}.}%
\end{figure}%
\begin{figure} \centering
\includegraphics[width = 0.7\textwidth]{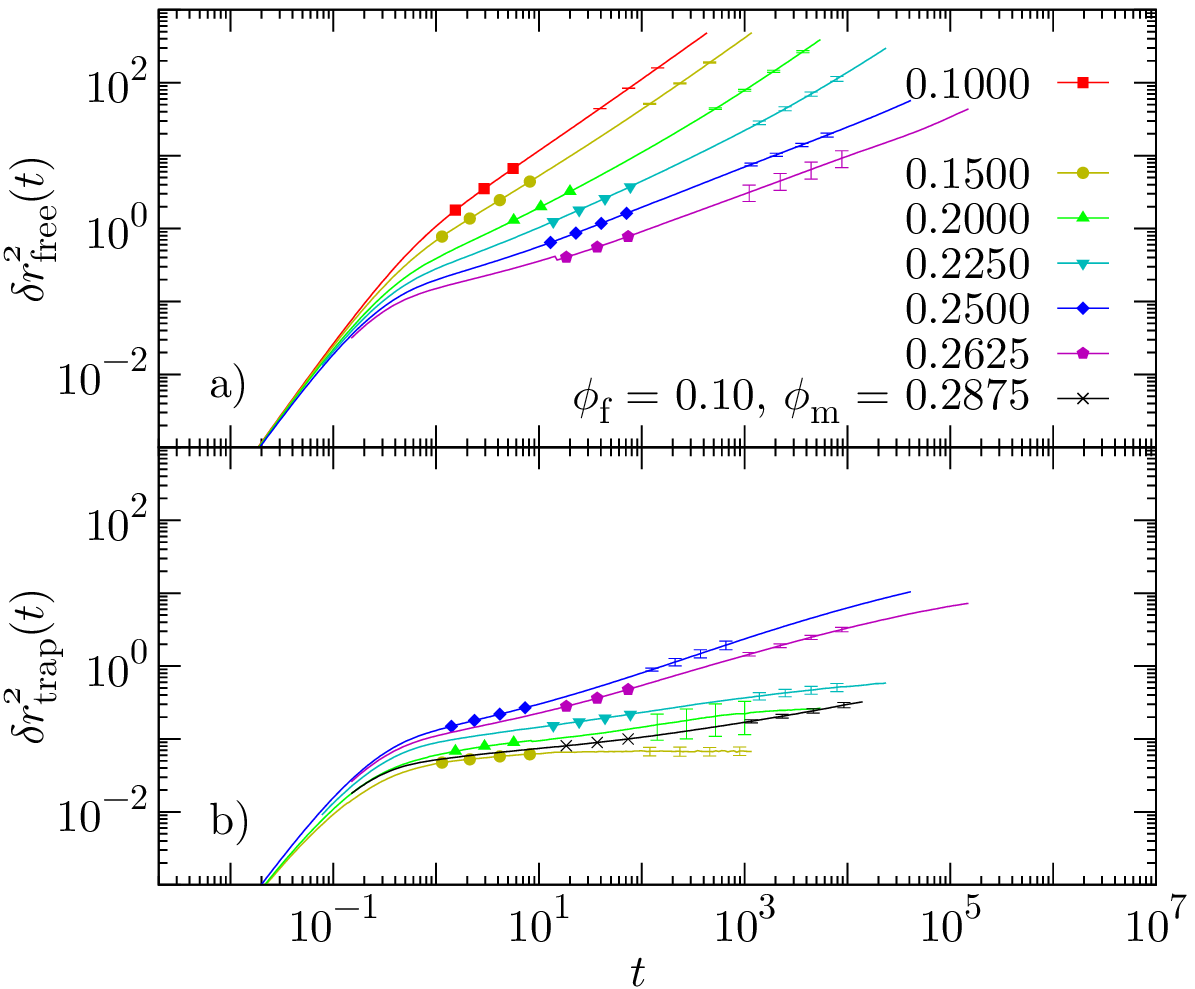}%
\caption{\label{fig:path3_msd}%
Mean-squared displacement as a function of time $t$ at fixed
$\Phif = 0.10$ for a series of $\Phim$. (a) for the free
particles, $\Drsf(t)$, (b) for the trapped particles, $\Drst(t)$.
Error bars: see Fig.~\ref{fig:path2_fs}.}%
\end{figure}%
\begin{figure} \centering
\includegraphics[width = 0.7\textwidth]{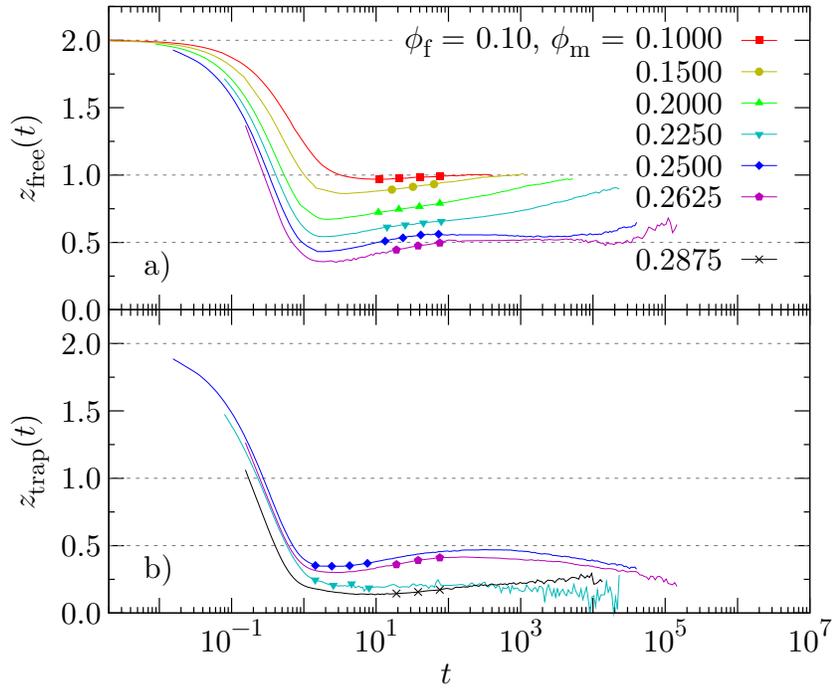}%
\caption{\label{fig:path3_z}%
Logarithmic derivative of the mean-squared displacement as a
function of time $t$ at fixed $\Phif = 0.10$ for a series of
$\Phim$. (a) for the free particles, $\Zf(t)$, (b) for the trapped
particles, $\Zt(t)$.}%
\end{figure}%

The contributions to $\Fs(k,t)$ along path~III are displayed in
Fig.~\ref{fig:path3_fs}. Disregarding in panel (a) the state point
above the percolation threshold (which is poorly equilibrated with
respect to the free particles), we find that $\Fsf(k{=}7,t)$ relaxes
all the way to zero for $t \rightarrow \infty$. This contrasts with
the single-particle intermediate scattering function of the full
fluid---\Cf\ Fig.~10(b) in Ref.~\cite{kurzidim2010}---which attains
a nonzero long-time plateau that increases with the matrix packing
fraction. Hence, the nondecaying part of $\Fs(k,t)$ must be entirely
due to the trapped particles; this fact is reflected in the
significant nonrelaxing part of $\Fst(k,t)$. We observe that $\Fst(k,
t{\rightarrow}\infty)$ exhibits a nonmonotonic behaviour, attaining a
minimum value at $\Phim \simeq 0.25$, which is close to the
percolation transition in the void space \cite{kurzidim2011a}.
$\Drsf(t)$ and $\Zf(t)$---the latter is presented in Fig.~\ref{fig:path3_z}---show
features similar to those of their equivalents
for the full fluid (\Cf\ Fig.~11 in \cite{kurzidim2010}), apart
from two differences: firstly, $\Zf(t)$ shows an onset to recover
diffusive behaviour for $\Phim=0.2625$, as opposed to $z(t)$ which
shows a trend to saturation. Therefore, in the time range in question
($t \simeq 10^5$) the mean-squared displacement of the full fluid has
to be dominated by the contribution of the trapped particles. The
second interesting feature is the much clearer subdiffusive regime for
$\Phim=0.2625$, which now remains at an essentially invariable $\Zf
\simeq 0.5$ for almost three decades in time. The mean-squared
displacement of the trapped particles reveals a similarly interesting
behaviour: while $\Drst(t)$, shown in Fig.~\ref{fig:path3_msd}, is
subdiffusive with $\Zt(t) < 0.5$ at all times, for matrix packing
fractions close to the percolation threshold $\Zt(t)$ attains much
larger values of almost $0.5$ at intermediate times, and $\Drst(t)$
saturates at much larger values in this $\Phim$ regime. The latter
indicates that the average trap size attains a maximum at $\Phim^*$.
This behaviour is well known from percolation theory, which predicts
the average trap size to diverge; onsets of this phenomenon were also
shown to exist in Refs. \cite{kurzidim2011a, sung2008}.

\begin{figure} \centering
\includegraphics[width = 0.7\textwidth]{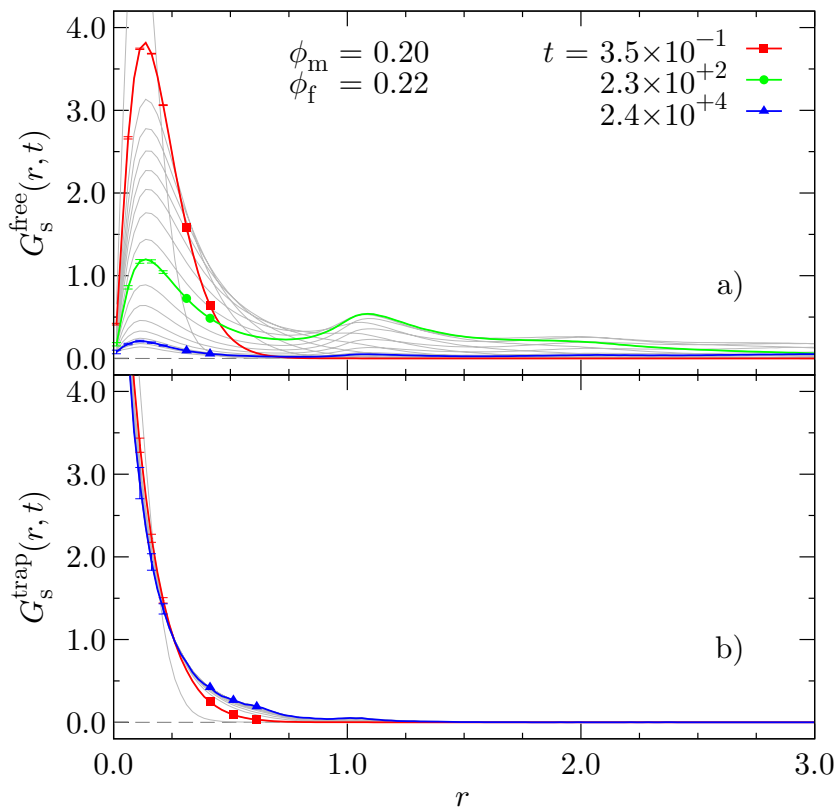}%
\caption{\label{fig:grt_upper-left}%
Self-part of the direction-averaged van Hove function as a
function of $r$ at point~``B'', defined via $\Phim = 0.20$ and
$\Phif = 0.22$. (a) for the free particles, $\Gsf(r,t)$, (b) for
the trapped particles, $\Gst(r,t)$. Curves are approximately
logarithmically spaced in time; some are highlighted for visual
guidance. Error bars: see Fig.~\ref{fig:path2_fs}.}%
\end{figure}%
\begin{figure} \centering
\includegraphics[width = 0.4\textwidth]{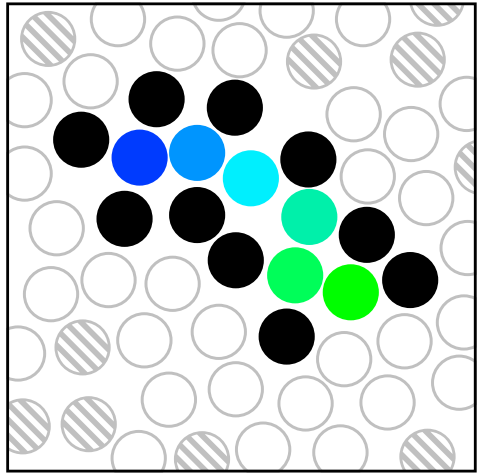}%
\caption{\label{fig:dead-end}%
Two-dimensional schematic of a long dead-end channel. Black solid
circles: matrix particles that form the dead-end channel. Grey
(coloured) solid circles: fluid particles in the channel. Light
grey hatched circles: other matrix particles. Light grey open
circles: other fluid particles. From light green to deep blue
(colour online) there is a decreasing probability for the particle
to leave its position via collective rearrangements.}%
\end{figure}%

The self-part of the direction-averaged van Hove function, $\Gs(r,t)$,
shall be discussed for two selected state points, labelled point~``B''
and point~``C'', which are indicated in Fig.~\ref{fig:kin_diagr}.
Point~B is located on path~II (fixed $\Phim = 0.20$) at a fluid
packing fraction $\Phif = 0.22$, where the latter value has been shown
sufficient to cause a relatively slow relaxation of the fluid's
dynamic features \cite{kurzidim2009, kurzidim2010}. The respective
contributions to the van Hove functions are depicted in Fig.~\ref{fig:grt_upper-left}.
The main peak of $\Gsf(r,t)$ at small distances
drops rapidly at small times. However, for large times it is still
present; this is probably due to fluid particles that are located at
the end of long, narrow dead-end channels like the one schematically
drawn in Fig.~\ref{fig:dead-end}. Those channels are connected to
the percolating void, but very improbable collective rearrangements
would have to occur for the end-particle to leave its position. At
intermediate times and $r \sim 1.1$ we see a pronounced peak in
$\Gsf(r,t)$, which indicates hopping processes of the free particles.
$\Gst(r,t)$, on the other hand, quickly reaches a stationary
distribution; the latter is relatively narrow which indicates that for
$\Phim=0.20$ only a small fraction of pores is large enough to
accommodate more than one particle. For large times the van Hove
function displays a small shoulder at $r \sim 0.5$, reflecting the
pore size distribution; the small peak at $r \sim 1$ emerging for
large times evidences that despite the small average trap size there
exist traps large enough for particles to collectively rearrange.

\begin{figure} \centering
\includegraphics[width = 0.7\textwidth]{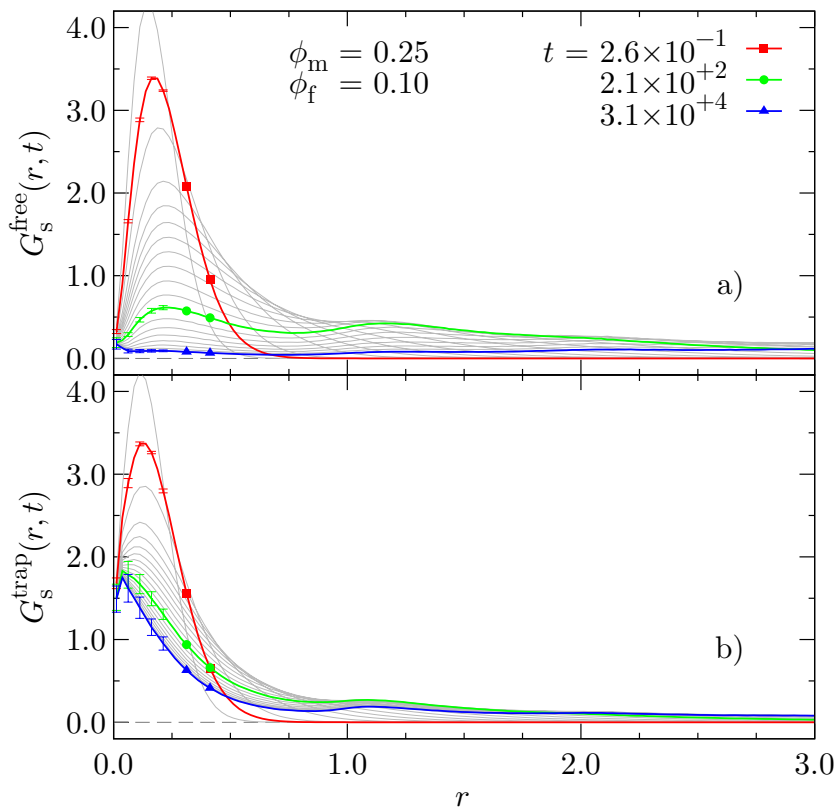}%
\caption{\label{fig:grt_lower-right}%
Self-part of the direction-averaged van Hove function as a
function of $r$ at point~``C'', defined via $\Phim = 0.25$ and
$\Phif = 0.10$. (a) for the free particles, $\Gsf(r,t)$, (b) for
the trapped particles, $\Gst(r,t)$. Curves are approximately
logarithmically spaced in time; some are highlighted for visual
guidance. Error bars: see Fig.~\ref{fig:path2_fs}.}%
\end{figure}%

Point~C, the other point at which we examined $\Gs(r,t)$, is located
on path~III and is characterised by $\Phif = 0.1$ and $\Phim = 0.25$;
notably, the latter figure is close to the percolation threshold. The
respective contributions to the van Hove functions are depicted in
Fig.~\ref{fig:grt_lower-right}. The main peak of $\Gsf(r,t)$ at $r
\sim 0$ drops to zero as $t$ tends to infinity. For $r = 0$ a minute
peak seems to persist; it might be associated with the trapping
phenomena discussed in Sec.~\ref{sec:Discussion}. Other than a
hopping peak at $r \sim 1$ (less pronounced than that in Fig.~\ref{fig:grt_upper-left})
there are no unusual features in $\Gsf(r,t)$. In
contrast, for this state point $\Gst(r,t)$ shows a behaviour which is
distinctively different from that of the same correlator for the other
state point: at point~C the evolution of the stationary state at $t
\rightarrow \infty$ takes considerably more time than at point~B (\Cf\
Fig.~\ref{fig:grt_upper-left}). Also, the pronounced peak at small
distances in the limiting function evidences a considerable amount of
trapping; the broad side peak at $r \sim 1.1$ reflects hopping
processes inside the traps. Finally, it is worth noting that
$\Gst(r,t{\rightarrow}\infty)$ exhibits a very long tail in $r$; this
is another indication of the fact that close to the percolation
transition there exist very large traps.

\section{Discussion}%
\label{sec:Discussion}%

Our analysis of the dynamic correlation functions for the free and the
trapped particles provides some deeper insight into the complex
relaxation of a colloidal fluid adsorbed in a dense immobile matrix.
Comparison with results for the full fluid \cite{kurzidim2009,
kurzidim2010} strengthens our previous interpretation of the dynamic
data and allows to assess the various predictions of MCT \cite{krakoviack2005,
krakoviack2007, krakoviack2009} from a broader
perspective:

(i) The complex relaxation of the single-particle correlators along a
pathway representative of the behaviour at intermediate matrix and
large fluid densities (path II) arises from a genuine superposition of
trapping effects and a collective caging mechanism: the former leads
to a finite long time plateau in $\Fs(k,t)$, while the latter is
responsible for the inflection in the intermediate time regime of this
correlation function. This observation strongly supports our previous
interpretation for the data of the the full fluid along this path
\cite{kurzidim2009, kurzidim2010} and indicates that this behaviour
may correspond, at least on a qualitative level, to the crossing of
the two arrest lines (diffusion-localisation and glass-transition) as
predicted by MCT.

(ii) The values of the subdiffusion exponent $z$ for the full fluid
\cite{kurzidim2009, kurzidim2010} clearly show that the free fluid
particles undergo a subdiffusive process on their own right. Moreover,
the values of $z$ for the full fluid are only marginally affected by
the contribution of the trapped particles: our calculation of $\Zf$ at
a packing fraction close to the estimated percolation transition is,
within error margin, indistinguishable from the value $0.5$ found in
the full system. As an observational fact, the same value of $z$ is
featured in single-file diffusion \cite{kaerger1992, wei2000} and
is also predicted by MCT along the diffusion-localisation transition
\cite{krakoviack2009}. However, it is well-known that MCT cannot
properly describe the diffusion of a tracer particle in a disordered
frozen matrix such as in the Lorentz gas model \cite{krakoviack2009,
schnyder2011}. In particular, in the Lorentz gas the subdiffusion
exponent $z = 0.32$ is observed at the percolation transition \cite{hoefling2006,
hoefling2007}, which differs from the value $z = 0.5$
predicted by MCT at the diffusion-localisation transition in QA
systems. Motivated by this observation and by related findings in
previous works \cite{kim2009, voigtmann2009}, we have examined the
case of non-interacting tracer particles moving in a QA-type matrix at
the percolation transition \cite{kurzidim_inpreparation}. We found
that in this case $z$ for the full fluid assumes a value around $0.3$,
which is more similar to the value found in the Lorentz gas than to
the value predicted by MCT. The apparent disagreement between the case
of noninteracting tracers and the case of a dense fluid moving in such
matrices will be subject to future work \cite{kurzidim_inpreparation}.

(iii) The splitting procedure effectively removes the nonzero
long-time limit of $\Fs(k,t)$. This method therefore provides a
microscopic basis to the empirical procedure described in \cite{kurzidim2010},
where an ad-hoc subtraction of the long time plateau
was performed. The agreement between the results of the two procedures
(not discussed here) lends support to our discussion of the modified
kinetic diagram for single-particle properties presented in \cite{kurzidim2010}.

\begin{figure} \centering
\begin{tabular}{c @{\hspace{1.0cm}} c}%
\includegraphics[width = 0.4\textwidth]{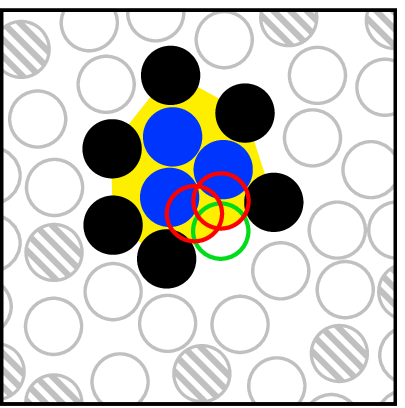} &
\includegraphics[width = 0.4\textwidth]{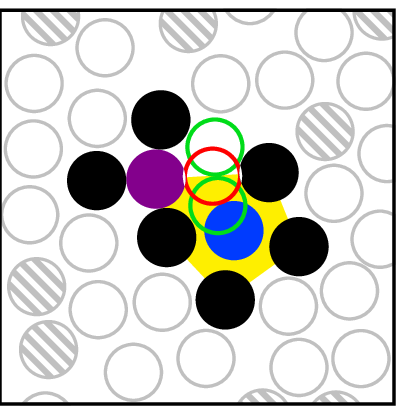}
\\
(a) & (b)%
\end{tabular}%
\caption{\label{fig:effective_trapping}%
Two-dimensional schematics for mechanisms of effective trapping,
induced by the presence of multiple fluid particles in the system.
Black solid circles: relevant matrix particles. Dark purple solid
circles: trapped particles. Dark blue solid circles:
effectively-trapped particles. Green open circles: positions a
fluid particle can occupy. Red open circles: positions a fluid
particle can \textit{not} occupy due to overlaps. Yellow: area
constituting the effective trap. Light grey hatched circles: other
matrix particles. Light grey open circles: other fluid particles
populating the percolating void. (a) The effectively-trapped
particles are located in a pore connected to the percolating void;
however, due to jamming the particles cannot rearrange such that
one of them would reach the connecting channel. (b) The
effectively-trapped particles are located in a pore connected to
the percolating void; however, another fluid particle (purple) is
trapped between matrix particles in a way that it blocks the
connecting channel for the particles in the pore (blue).}%
\end{figure}

Even more intriguing is the observation of nontrivial dynamic
heterogeneity in both the trapped and the free component of the fluid
particles. For the free particles, the van Hove function displayed in
Fig.~\ref{fig:grt_lower-right}(a) evidences that a non-negligible
fraction of the particles undergo a very slow relaxation. At least
three physical mechanisms explaining this finding can be identified.
(i) Free particles may be located at the end of long dead-end channels
as shown in Fig.~\ref{fig:dead-end}; the longer such a channel is,
the less likely are the collective rearrangements required for a
particle to escape from the channel's end. (ii) Free particles may
mutually jam in a pore as schematised in Fig.~\ref{fig:effective_trapping}(a);
such particles are \textit{rigorously}
confined to that pore since none of them can reach the channel that
connects the pore to the percolating void. (iii) A trapped particle
may act as an ``effective'' matrix particle [see Fig.~\ref{fig:effective_trapping}(b)]:
if the matrix restricts its movement to
much less than its radius, it can block a pathway in the percolating
void; this may render the void at one side of that pathway a trap.
Visual inspection of animated trajectories of our systems confirmed
that all of these processes are present and relevant in the portion of
the kinetic diagram corresponding to large $\Phif$ and $\Phim$. Note
that effects (ii) and (iii) arise from the QA-protocol requirement
that all available volume (including traps) be populated with fluid
particles.

Notably, dynamic heterogeneity is also present among the trapped
particles; they have been observed to rearrange collectively for
instance by exchanging their positions in a closed loop. Thus, for the
trapped particles an apparent weak relaxation appears in the
single-particle correlators and a tail emerges at large distances in
their van Hove function. A more systematic analysis of these striking
dynamic features is currently underway.

\section{Conclusions}%
\label{sec:Conclusions}%

We have presented a computational study concerned with the dynamic
properties of a model colloidal fluid adsorbed in a disordered, rigid
array of obstacles. The original aspect of this study consists in the
splitting of dynamic correlators into two contributions, one
originating from the free and the other from the trapped fluid
particles. The notion of ``free'' and ``trapped'' refers to the
structure of the voids formed by the matrix: free particles are
located in the infinitely-large percolating void whereas trapped
particles reside in disconnected voids. We compared the relaxation
patterns of the two components with the corresponding patterns for the
full fluid, which allows to disentangle the dynamic effects due of
crowding and confinement. Further, our procedure provides some deeper
insight into the relaxation processes at play at the microscopic
level. Our analysis confirmed the striking superposition of two
relaxation mechanisms---trapping and caging---at a sufficiently large
density of both the fluid and the matrix component, and supported the
view that MCT correctly captures the qualitative features of the
dynamics in the corresponding portion of the kinetic diagram. Our
numerical procedure also unveils that an unexpected dynamic
heterogeneity is present in the motion of fluid particles within the
matrix structure: on the one hand, even particles residing in a
percolating void may be effectively trapped; on the other hand,
residual relaxation can take place within disconnected pores. We
believe these aspects to be relevant also for the assessment of
dynamic heterogeneity in more general models of fluids adsorbed in
porous media. Whether these complex dynamic phenomena are present also
in systems with different protocols to generate the matrix \cite{kim2009,
krakoviack2010a}, soft interactions \cite{kurzidim2009},
or systems in which a large dynamic asymmetry replaces the quenched
disorder \cite{moreno2006, fenz2009, voigtmann2009} is a question
that remains to be addressed in future studies.

\ack We acknowledge fruitful discussions with V.~Krakoviack and
T.~Franosch. This work was financially supported by the Austrian
Research Fund (FWF) under Proj. Nos. P19890-N16 and W004.

\section* {References}%

\end{document}